\documentclass[aps,preprintnumbers,amsmath,amssymb,nofootinbib,superscriptaddress,notitlepage,10pt]{revtex4-1}

\usepackage{comment}
\usepackage{amssymb}
\usepackage{color}
\usepackage{rotating}
\usepackage{amsmath}
\usepackage{ulem}
\usepackage{overpic}
\usepackage{graphicx}
\usepackage{dcolumn}
\usepackage{graphicx}
\usepackage{bm}
\usepackage{chngpage}
\usepackage{multirow}
\usepackage{slashed}
\usepackage{indentfirst}
\usepackage{booktabs}
\usepackage{amssymb,amsfonts}
\usepackage{amsmath}
\usepackage{color}
\usepackage{hyperref}

\begin{document}

\title{Charge radii of  potassium isotopes in the RMF(BCS)* approach}
\author{Rong An}

\affiliation{Key Laboratory of Beam Technology of Ministry of Education, Institute of Radiation Technology, Beijing Academy of Science and Technology, Beijing 100875, China}
\affiliation{Key Laboratory of Beam Technology of Ministry of Education, College of Nuclear Science and Technology, Beijing Normal University, Beijing 100875, China}
\author{Shi-Sheng Zhang}
\email[E-mail: ]{zss76@buaa.edu.cn}
\affiliation{School of Physics, Beihang University, Beijing 100191, China}

\author{Li-Sheng Geng}
\email[E-mail: ]{lisheng.geng@buaa.edu.cn}
\affiliation{School of Physics, Beihang University, Beijing 100191, China}
\affiliation{Beijing Key Laboratory of Advanced Nuclear Materials and Physics,  Beihang University, Beijing 102206, China}
\affiliation{School of Physics and Microelectronics, Zhengzhou University, Zhengzhou, Henan 450001, China}

\author{Feng-Shou Zhang}
\email[E-mail: ]{fszhang@bnu.edu.cn}
\affiliation{Key Laboratory of Beam Technology of Ministry of Education, Beijing Radiation Center, Beijing 100875, China}
\affiliation{Key Laboratory of Beam Technology of Ministry of Education, College of Nuclear Science and Technology, Beijing Normal University, Beijing 100875, China}
\affiliation{Center of Theoretical Nuclear Physics, National Laboratory of Heavy Ion Accelerator of Lanzhou, Lanzhou 730000, China}


\begin{abstract}
  We apply the recently proposed RMF (BCS)* ansatz to study the charge radii of the potassium isotopic chain up to $^{52}$K. It is shown that the experimental data can be reproduced rather well, qualitatively similar to the Fayans nuclear density functional theory, but with a slightly better description of the odd-even staggerings (OES). Nonetheless, both methods fail for $^{50}$K and to a lesser extent for $^{48,52}$K. It is shown that if these nuclei are deformed with a $\beta_{20}\approx-0.2$, then one can obtain results consistent with experiments for both charge radii and spin-parities. We argue that beyond-mean-field studies are needed to properly describe the charge radii of these three nuclei, particularly for $^{50}$K.
\end{abstract}


\maketitle
\section{INTRODUCTION}
 Charge radii are fundamental quantities that describe atomic nuclei. A rule of thumb is  that  they scale with either masses  as $A^{1/3}$~\cite{P.Ring,A.Bohr} or charges as $Z^{1/3}$~\cite{Zhang:2001nt,KNWL200203009}. With the rapid development of novel detectors and analysis techniques,  charge radii of many atomic nuclei far away from the $\beta$-stability line have been measured with high precision, such as calcium~\cite{Ruiz2016,Miller2019}, cadmium~\cite{Hammen2018}, tin~\cite{Gorges2019},  mercury~\cite{goodacre2020laser}, copper~\cite{deGroote}, and potassium isotopes~\cite{PhysRevC.100.034304,Koszorus2020mgn}. However, for certain nuclei, large discrepancies are observed between experimental measurements and theoretical predictions. For instance, a parabolic-like shape and strong odd-even staggering (OES) effects have long been known to exist in the calcium isotopes  between $^{40}$Ca and $^{48}$Ca~\cite{ANGELI201369}. Such peculiar features persist toward the neutron-deficient region~\cite{Miller2019}. Beyond $N=29$, the charge radii increase rapidly and the radius of  $^{52}$Ca is significantly larger than that of $^{48}$Ca~\cite{Ruiz2016}. This is unexpected because $N=32$ was believed to be a magic number in the calcium isotopes~\cite{nature498,PhysRevC.31.2226}.

In comparison with the charge radii of calcium isotopes, the amplitude of the parabolic-like shape between $^{39}$K and $^{47}$K is smaller due to the last unpaired proton~\cite{ANGELI201369,TOUCHARD1982169,Martensson_Pendrill_1990,Bendali_1981,PhysRevA.74.032503}. Meanwhile, the rapid increase of charge radii is also found across the $N=28$ shell closure~\cite{KREIM201497}. The neutron-rich shell closure  at $N=32$ in the potassium isotopic chain was investigated in Refs.~\cite{PhysRevC.31.2226,PhysRevLett.114.202501}, and shows relatively enhanced stability. Recently, the collinear resonance ionization spectroscopy (CRIS) technique has been employed to measure the charge radii of potassium isotopes~\cite{PhysRevC.100.034304}, and the precision measurement of charge radii beyond $N=32$ has been performed for the first time below $Z<20$ for potassium isotopes~\cite{Koszorus2020mgn}. No sudden increase of the charge radius of $^{52}$K was observed.

All of these results challenge our understanding of the evolution of nuclear charge radii of exotic isotopes with large neutron or proton excesses. To address these challenges, many novel approaches have been proposed. In Ref.~\cite{RenZZ2020}, a statistical method is introduced to study nuclear charge radii by combining  sophisticated nuclear models with the naive Bayesian probability (NBP) classifier. This method predicts a rapid increase of charge radii beyond $N=28$. In Ref.~\cite{Wu:2020bao}, a feed-forward neural network model which relates charge radii to the symmetry energy is explored. The strong increase in the charge radii beyond $N=28$ is well reproduced by the Fayans energy density functional (EDF) model~\cite{Koszorus2020mgn,PhysRevC.95.064328}. However, this method overestimates the OES effect of the charge radii of the potassium isotopic chain. In addition,  the deviation between experiment and theory becomes larger toward the neutron-deficient region~\cite{Koszorus2020mgn}.
In Ref.~\cite{An:2020qgp}, we proposed an empirical ansatz based on the relativistic mean field (RMF) theory, which adds a correction term induced by the difference of pairing interactions for protons and neutrons calculated self-consistently in the RMF. This modified approach can remarkably reproduce the OES effects of charge radii of calcium isotopes and nine other even-Z isotopic chains, especially the strong increase of  charge radii across the $N=28$ shell closure along the calcium isotopic chain. In this work, we would like to extend the ansatz of Ref.~\cite{An:2020qgp} to study the root mean square (rms) charge radii of odd-proton potassium isotopes.

This work is organized as follows. In Sec. II,  the theoretical framework is briefly introduced. The results and corresponding discussions are presented in Sec. III. In the last section, we present the conclusions.

\section{Theoretical Framework}
In the past three decades, relativistic mean field (RMF) theories have achieved remarkable successes in describing properties of finite nuclei around and far away from the $\beta$-stability line~\cite{PhysRevLett.77.3963,MENG19983,PhysRevC.68.034323,VRETENAR2005101,PhysRevC.82.011301,PhysRevC.85.024312,Liang:2014dma,jie2016relativistic,SUN2018530,2020sun,PhysRevC.102.024314,PhysRevC.104.L021301}, not only for ground states but also for excited states~\cite{PhysRevC.67.034312,PhysRevC.69.054303,zhang2007,PhysRevC.90.044305,zhang2012,PhysRevC.92.024324,Cao:2003yn}. In this work, we adopt the meson-exchange version of RMF. The nonlinear Lagrangian density, where nucleons are described as Dirac particles and interact via the exchange of $\sigma$, $\omega$ and $\rho$ mesons, has the following form:
\begin{eqnarray}
\mathcal{L}&=&\bar{\psi}[i\gamma^\mu\partial_\mu-M-g_\sigma\sigma
-\gamma^\mu(g_\omega\omega_\mu+g_\rho\vec
{\tau}\cdotp\vec{\rho}_{\mu}+eA_\mu)]\psi\nonumber\\
&&+\frac{1}{2}\partial^\mu\sigma\partial_\mu\sigma-\frac{1}{2}m_\sigma^2\sigma^2
-\frac{1}{3}g_{2}\sigma^{3}-\frac{1}{4}g_{3}\sigma^{4}\nonumber\\
&&-\frac{1}{4}\Omega^{\mu\nu}\Omega_{\mu\nu}+\frac{1}{2}m_{\omega}^2\omega_\mu\omega^\mu
+\frac{1}{4}c_{3}(\omega^{\mu}\omega_{\mu})^{2}\nonumber\\
&&-\frac{1}{4}\vec{R}_{\mu\nu}\cdotp\vec{R}^{\mu\nu}+\frac{1}{2}m_\rho^2\vec{\rho}^\mu\cdotp\vec{\rho}_\mu
+\frac{1}{4}d_{3}(\vec{\rho}^{\mu}\vec{\rho}_{\mu})^{2}\nonumber\\
&&-\frac{1}{4}F^{\mu\nu}F_{\mu\nu}.
\end{eqnarray}
Where $M$ is the nucleon mass and $m_{\sigma}$, $m_{\omega}$, and $m_{\rho}$ are the masses of the $\sigma$, $\omega$, and $\rho$ mesons; $A_\mu$ is the photon field and $F^{\mu\nu}$ is the electromagnetic tensor; and $g_{\sigma}$, $g_{\omega}$, $g_{\rho}$, and $e^{2}/4\pi$ are the coupling constants for the $\sigma$, $\omega$, $\rho$ mesons, and photon, respectively. For the mean-field parameters, we choose the NL3 parameter set~\cite{PhysRevC.55.540}. The Dirac equation for the nucleons and the Klein-Gordon-type equations with sources for the mesons and the photon are solved by the expansion method with the axially symmetric harmonic oscillator basis~\cite{Ring:1997tc,Geng:2003pk}. We use $12$ shells for expanding the fermion fields and $20$ shells for the meson fields.

In the conventional RMF(BCS) model, the mean square charge radius is calculated in the following way (in units of fm$^{2}$)~\cite{Ring:1997tc,Geng:2003pk}:
\begin{eqnarray}\label{ch1}
\langle{R_{\mathrm{ch}}^{(2)}}\rangle=\frac{\int{r}^{2}\rho_{p}(r)d^{3}r}{\int\rho_{p}(r)d^{3}r}+0.64~\mathrm{fm}^2,
\end{eqnarray}
where the first term represents the charge distribution of point-like protons and the second term is from the finite size of protons~\cite{Ring:1997tc}. To account for the experimentally-observed odd-even staggerings of charge radii, Ref.~\cite{An:2020qgp} proposed a modified formula:
\begin{eqnarray}\label{coop1}
\langle{R_{\mathrm{ch}}^{(2)}}\rangle=\frac{\int{r}^{2}\rho_{p}(r)d^{3}r}{\int\rho_{p}(r)d^{3}r}+0.64~\mathrm{fm}^2+\frac{a_{0}}{\sqrt{A}}\Delta{\mathcal{D}}~\mathrm{fm}^2.
\end{eqnarray}
The last term on the right hand is the correction term that  can be associated to Cooper pair condensations~\cite{PhysRevC.76.011302}. The quantity $A$ is the mass number and $a_{0}=0.834$ is a normalization constant obtained  by fitting to the  experimental charge radii. The quantity $\Delta\mathcal{D}=|\mathcal{D}_{n}-\mathcal{D}_{p}|$ represents the difference of Cooper pair condensations for neutrons and protons. It is calculated self-consistently by solving the state-dependent BCS equations with a $\delta$ force~\cite{Geng:2003pk,20181107}.
One should note that although the correction is introduced as an empirical approximation of neutron-proton pairing correlations, it is calculated using the outputs of the microscopic RMF(BCS) approach. For the convenience of discussion, the results obtained by Eq.~(\ref{ch1}) are labeled as RMF(BCS), and RMF(BCS)* represents the calculated results though the modified charge radius formula Eq.~(\ref{coop1}). More discussions can be found in Ref.~\cite{An:2020qgp}.

\section{RESULTS AND DISCUSSIONS}
\subsection{Charge radii of  potassium isotopes}

In the RMF(BCS)* approach, the strength of the  pairing interaction is determined by fitting to the odd-even staggerings of binding energies. For this purpose, the following three-point formula is employed~\cite{P.Ring,A.Bohr}:
\begin{eqnarray}
\Delta_{E}=\frac{1}{2}[B(N-1,Z)-2B(N,Z)+B(N+1,Z)],
\end{eqnarray}
where $B(N,Z)$ is the binding energy for a nucleus of neutron number $N$ and proton number $Z$. In this study, the pairing strength is fixed at $350$~MeV~fm$^{3}$ and the pairing space is chosen to include all the single particle levels  within  24 MeV above and below the Fermi surface. To study nuclei with an odd number of nucleons, a  blocking approximation is adopted. At each step of the self-consistent iteration in solving the RMF equations, the last single particle level occupied by the odd nucleon is blocked~\cite{Geng:2003wt}. One should note that in certain cases, e.g., where the single particle levels around the Fermi surface are dense or configuration mixing is  particularly relevant, such a procedure may not  yield accurate results.

\begin{figure}
\centering
\includegraphics[width=0.6\linewidth]{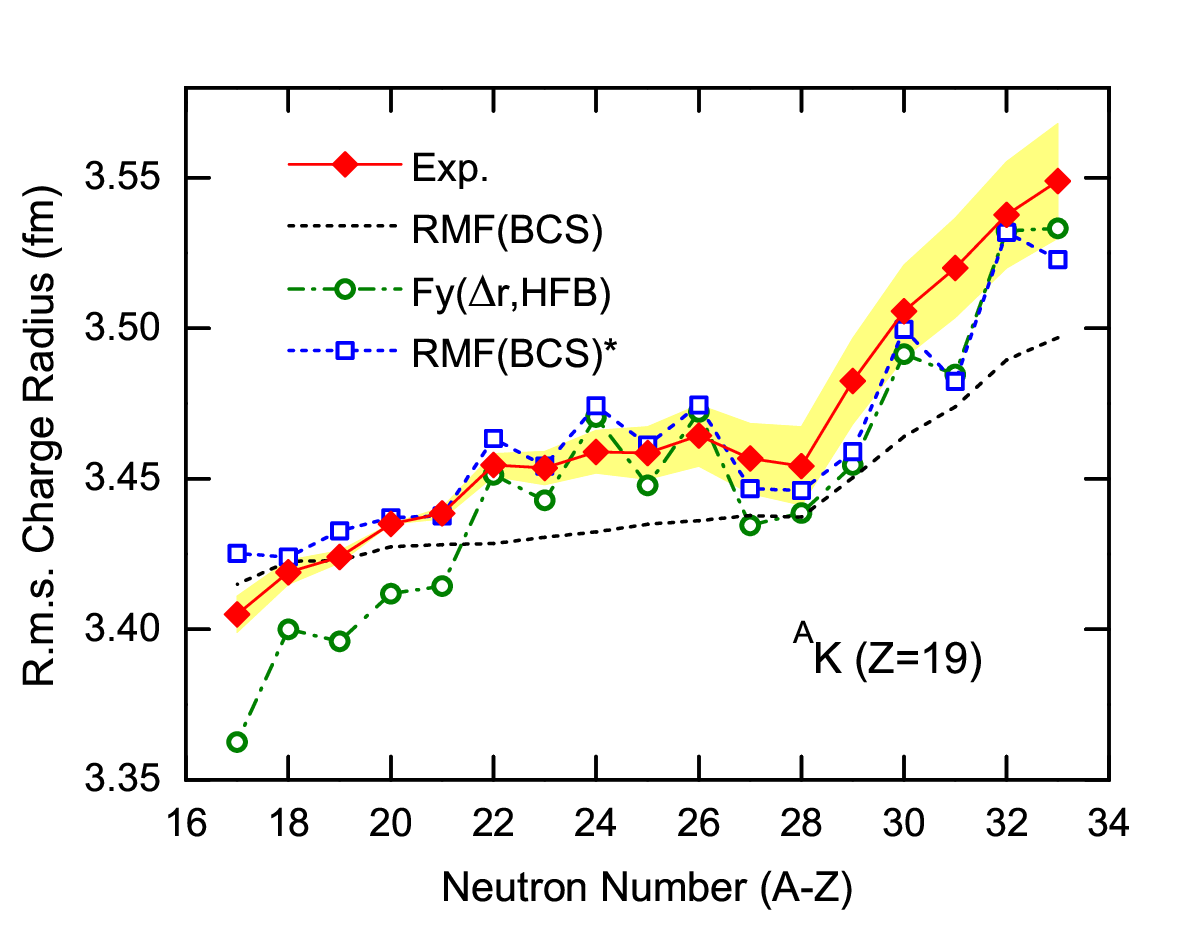}
\caption{Charge radii of potassium isotopes obtained in the RMF(BCS) method and  RMF(BCS)* ansatz. The experimental data are taken from Refs.~\cite{ANGELI201369,Koszorus2020mgn} and the yellow band indicates the systematic error. The Fayans EDF results~\cite{Koszorus2020mgn} are also shown for comparison.}\label{fig1}
\end{figure}
In Ref.~\cite{Koszorus2020mgn}, the charge radius of the exotic $^{52}$K isotope beyond the $N=32$  shell closure was measured. Similar to the calcium isotopes, a parabolic-like shape is also found for the potassium  isotopes between $^{39}$K and $^{47}$K~\cite{ANGELI201369}. However, the OES effects are much reduced. In Fig.~\ref{fig1}, we compare the charge radii of potassium isotopes calculated in the RMF(BCS) method with and without the correction term. It is clear that the RMF(BCS) approach cannot describe the charge radii of potassium isotopes, particularly the strong increase of charge radii  beyond $N=29$ and the OES behaviors. On the other hand, the new RMF(BCS)* ansatz can reproduce the parabolic-like shape  between $^{39}$K and $^{47}$K.  In addition, the OES effects are also reproduced but  slightly overestimated. Compared with  the more sophisticated Fayans EDF model, the RMF(BCS)* results are in better agreement with data~\cite{Koszorus2020mgn}. Furthermore, we calculated quantitatively the standard root mean square (rms) deviation between $R_{\mathrm{expt.}}$ and $R_{\mathrm{theo.}}$ along the potassium isotopic chain. For the Fayans model, the standard rms deviation is $0.0214$ fm. By contrast, the rms deviation falls to $0.0150$ fm with the RMF(BCS)$^{*}$ approach. In particular, toward the neutron-deficient side, a large deviation can be found between experimental data and those of the Fayans EDF model, while the RMF(BCS)* approach yields results in better agreement with experiment. Beyond the $N=28$ shell closure, both the RMF(BCS)* approach and Fayans EDF model can reproduce the fast increase. However, both  underestimate the charge radii of $^{50}$K and $^{52}$K, particularly the former.

\subsection{Double odd-even staggering effects}
Similar to binding energies, one can also  define a three-point formula to extract OES for charge radii~\cite{PhysRevC.95.064328}
\begin{eqnarray}\label{eq1}
\Delta_{r}=\frac{1}{2}[R(N-1,Z)-2R(N,Z)+R(N+1,Z)],
\end{eqnarray}
where $R(N,Z)$ is the rms charge radius. It should be noted that the OES effects on charge radii, i.e., that nuclear charge radii of odd-neutron isotopes are smaller than the average of their even-neutron neighbours, have been observed throughout the nuclear chart~\cite{ANGELI201369}. Various possible explanations have been proposed, such as the blocking of ground state quadrupole vibrations by the odd neutron~\cite{REEHAL1971385}, core polarizations by valence neutrons~\cite{TALMI1984189,CAURIER198015}, $\alpha$ clusters~\cite{Zawischa:1985qds}, three- or four-body  residual interactions~\cite{ZAWISCHA1987299,PhysRevLett.61.149}, special deformation effects~\cite{GIROD19821,Ulm:1986wd}, neutron pairing energies~\cite{weber2005effects}, pairing correlation~\cite{PhysRevC.95.064328}, etc. Therefore, it is worthwhile checking whether this empirical ansatz can provide a reasonable description of the OES of charge radii of the odd-$Z$(=19) potassium isotopes.

\begin{figure}
\centering
\includegraphics[width=0.5\linewidth]{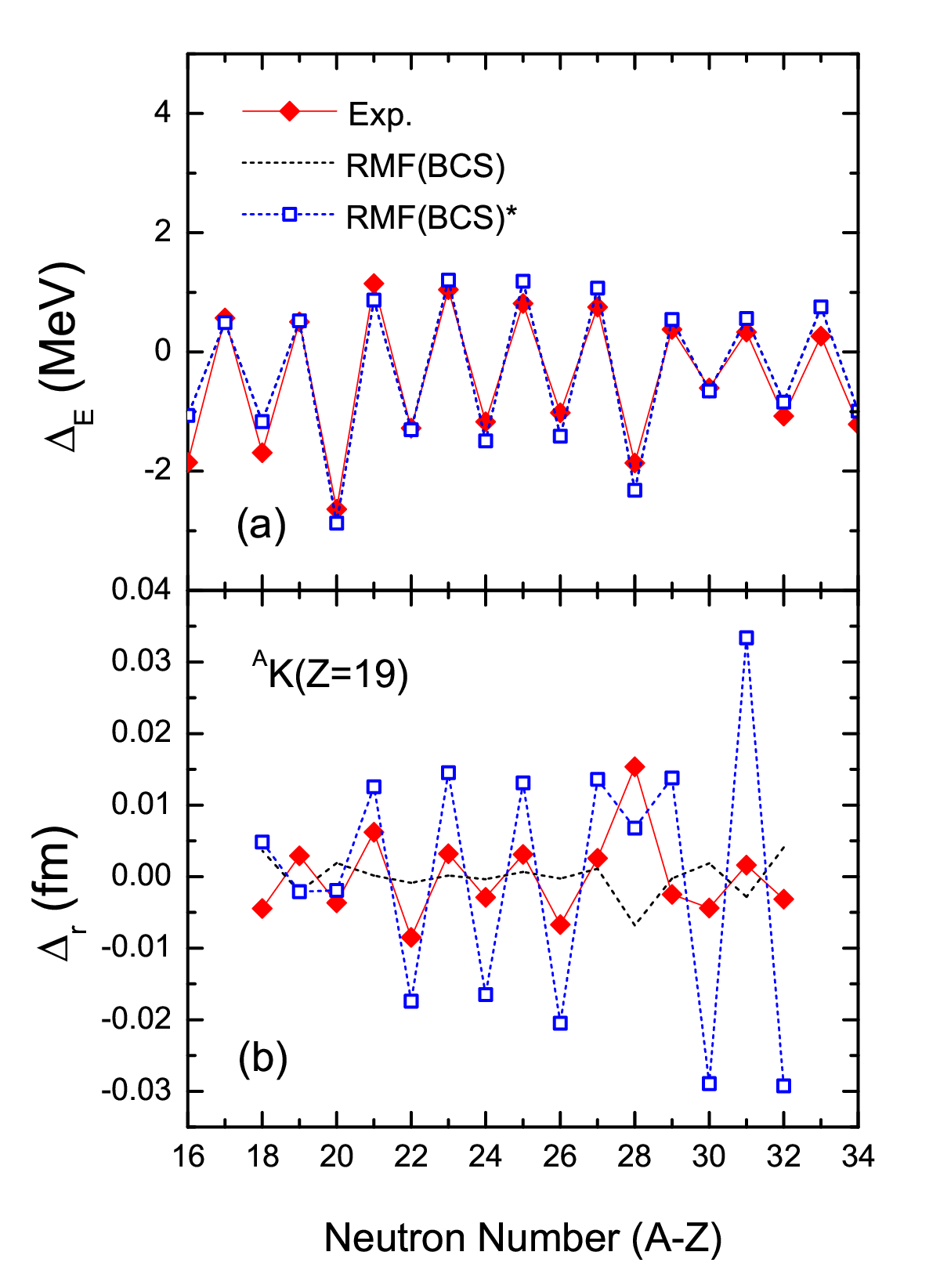}
\caption{ Odd-even staggerings of binding energies (a) and charge radii (b) of potassium isotopes. The experimental data for binding energies are taken from Ref.~\cite{Wangmeng30002}, while those of charge radii are from Refs.~\cite{ANGELI201369,Koszorus2020mgn}.}\label{fig2}
\end{figure}
In Fig.~\ref{fig2}, the OES of binding energies (upper panel) and charge radii (lower panel) are compared with the experimental data. The RMF(BCS) and RMF(BCS)* methods can reproduce the OES of binding energies rather well. For the potassium isotopic chain, the general trend of the OES of charge radii calculated by Eq.~(\ref{eq1}) is reproduced as well. However, the RMF(BCS)* method overestimates the OES of charge radii, especially in the neutron-rich region, as can be inferred already from Fig.~\ref{fig1}. For $^{37,38}$K, the OES behaviors are reversed due to the slightly overestimated charge radii of $^{36,38}$K.
For $^{47}$K, the modified formula cannot reproduce the local variation of the OES in charge radii. Meanwhile the general oscillation trend in experimental OES of charge radii is weakened at the $N=28$ shell closure, reversing the OES behavior. Actually, this phenomenon is also naturally observed at neutron magic numbers $N=50$, $82$, $126$~\cite{ANGELI201369}. In this work, the blocking approximation is employed to tackle the simultaneously unpaired proton and neutron. The overestimation of OES in nuclear charge radii may be corrected by tackling the last unpaired nucleons.

As one can see from Fig.~\ref{fig1}, the charge radii of $^{50}$K and $^{52}$K are underestimated, particularly, in both the RMF(BCS)* approach and the Fayans density functional theory. Such a deviation may be due to the blocking effect of unpaired nucleons, as mentioned above.
Along the isotopic chain, the added neutrons are mainly located in the outer edge of the nucleus for neutron-rich isotopes. As stressed in Ref.~\cite{Miller:2018mfb}, the neutron-proton (np) pairing correlation can cause protons to move closer to the added neutrons and increase the nuclear charge radius. Especially for the unpaired neutron and proton, the np-pairs could play a larger role. As a result, in the following, we will study the blocking treatment of the last unpaired proton/neutron.
\begin{figure}
\centering
\includegraphics[width=0.6\linewidth]{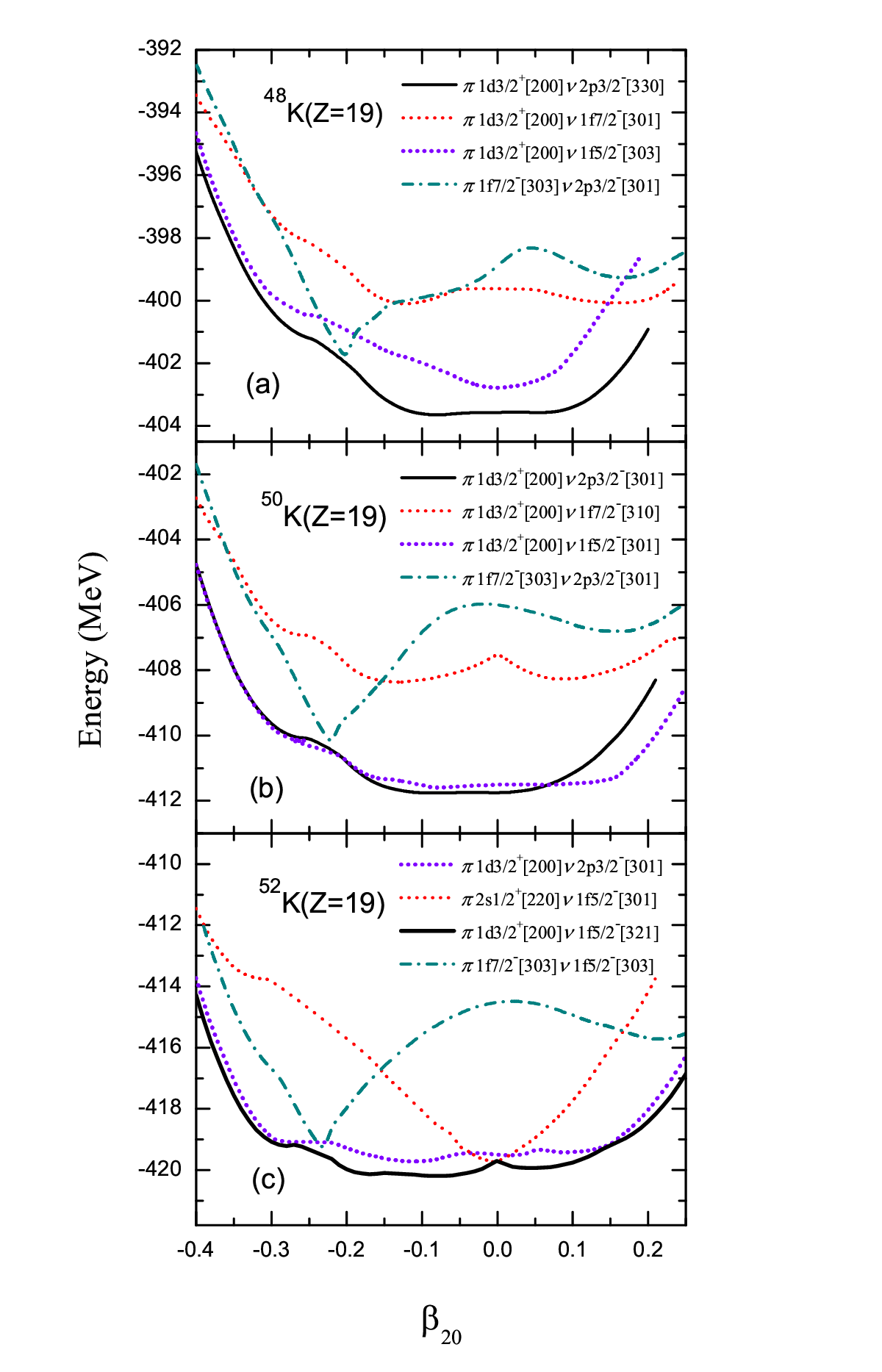}
\caption{Potential energy surfaces of $^{48}$K (a), $^{50}$K (b) and $^{52}$K (c) as functions of the quadrupole deformation parameter $\beta_{20}$ for different combinations of single particle orbits occupied by the last unpaired nucleons. $\pi$ and $\nu$ denote the last unpaired proton and neutron, where the combinations of spherical ($s, p, d, f$) and Nilsson quantum numbers in the square brackets are employed to denote the occupied orbits. }\label{fig3}
\end{figure}

\subsection{Blocking effects on charge radii}
\begin{figure}
\centering
\includegraphics[width=0.6\linewidth]{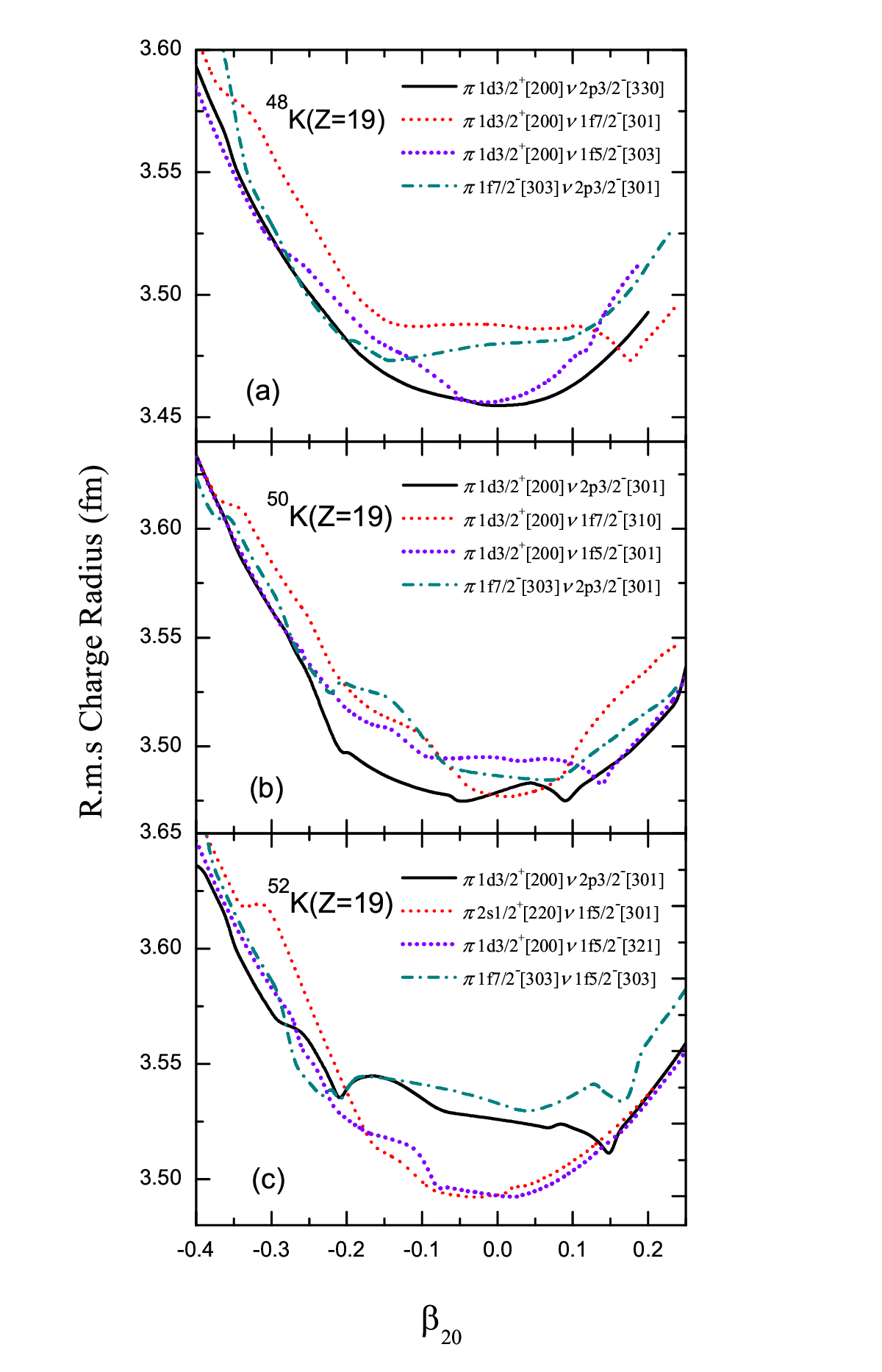}
\caption{ Same as Fig.~\ref{fig3} but for root mean square (rms) charge radii. }\label{fig6}
\end{figure}

In Figs.~\ref{fig3} and~\ref{fig6}, the potential energy surfaces and root mean square (rms) charge radii of $^{48}$K (a), $^{50}$K (b) and $^{52}$K (c) as a function of the quadrupole deformation parameter $\beta_{20}$ are plotted, with different assignments of single particle orbits occupied by the last unpaired proton ($\pi$) and neutron ($\nu$). The occupied orbits are given by the combinations of spherical ($s, p, d, f$) and Nilsson quantum numbers $[N,n_{z},m_{l}]$ in the square bracket. Where $N$ is the main quantum number and $n_{z}$ is the projection of $N$ on the z-axis, and $m_{l}$ is the component of the orbital angular momentum~\cite{Ring:1997tc}. Below, for the convenience of discussion, we use an expression such as ($1d_{3/2},2p_{3/2}$) to denote the occupied orbitals of the last unpaired proton (the first term in the  bracket) and neutron (the last term in the bracket).

In principle, the occupation of singe particle levels is determined self-consistently so that the largest binding energy is obtained. In such a way, the last unpaired proton is found to occupy the $1d_{3/2}$ orbital. The last unpaired neutron in $^{48}$K and $^{50}$K is found to occupy the  $2p_{3/2}$ orbital, while that in $^{52}$K occupies the $1f_{5/2}$ orbital. These configurations yield the largest binding energy. On the other hand, the potential energy surfaces are relatively soft. This implies that beyond-mean-field studies, which take into account configuration mixing, might be needed to correctly describe these nuclei.

For $^{48}$K, the configuration ($1d_{3/2},1f_{5/2}$) indicates a spherical ground state, but the corresponding charge radius is found to be much smaller than the  experimental radius. If the ($1d_{3/2},1f_{7/2}$) configuration is chosen,  the binding energy of $^{48}$K is smaller by 2 MeV and the charge radius is increased to  $3.487$ fm (see Fig.~\ref{fig6}). On the other hand, the configuration of ($1f_{7/2},2p_{3/2}$) leads to a smaller binding energy (in comparison with the $(1d_{3/2},2p_{3/2})$ configuration) and a charge radius of $3.481$ fm, which agrees with the experimental result. Therefore, for $N=29$, we conclude that either the last unpaired nucleons prefer to occupy the  ($1f_{7/2},2p_{3/2}$) states instead of the self-consistent ($1d_{3/2},2p_{3/2}$) configuration, or the nucleus is deformed with $\beta_{20}\approx-0.2$. Both scenarios cannot be realized in the present self-consistent calculation. Similar conclusions can be drawn for $^{50}$K.

For $^{52}$K, from the energy point of view, the most favored configuration is ($1d_{3/2},1f_{5/2}$). As one can see from Fig.~\ref{fig3}~(c), the differences between the four configurations are relatively small. Both ($1d_{3/2},2p_{3/2}$) and ($1f_{7/2},1f_{5/2}$) can yield results in reasonable agreement with data.

In Fig.~\ref{fig6}, one can find that the quadrupole deformation has an influence on rms charge radii of $^{48}$K (a), $^{50}$K (b) and $^{52}$K (c). Close to spherical shape, the rms charge radii are gradually reduced under various occupations. Toward larger deformation, especially around $\beta_{20}\approx-0.2$, comparable values are obtained. Combining the potential energy surfaces and the rms charge radii of $^{48}$K, $^{50}$K and $^{52}$K, the occupations of $1f_{7/2}$ levels with the last unpaired proton seem to give plausible values of the charge radii.

\begin{figure}
\centering
\includegraphics[width=0.6\linewidth]{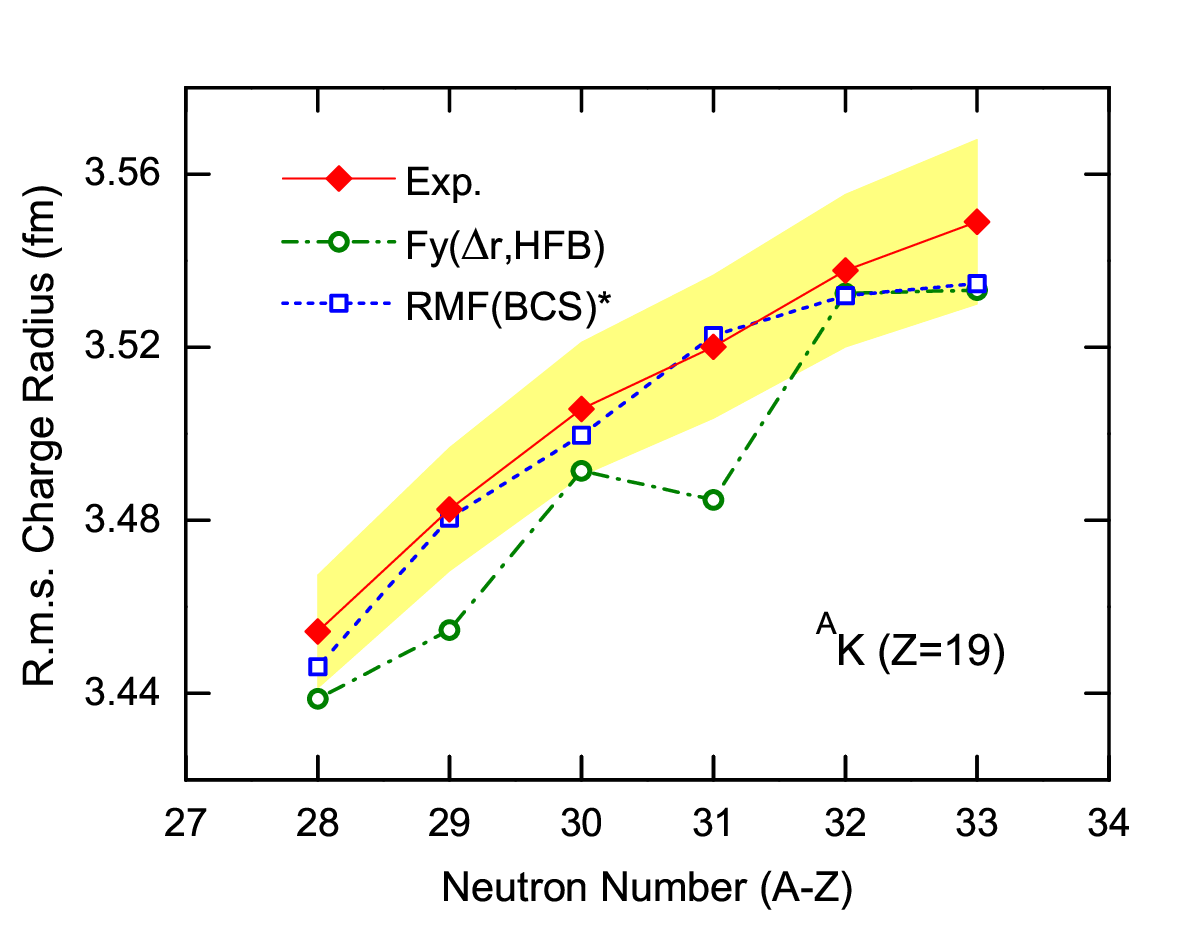}
\caption{Same as Fig.~\ref{fig1}, but with
different treatments of the blocking approximation for the last unpaired proton and neutron beyond $N=28$. }\label{fig4}
\end{figure}

In Fig.~\ref{fig4}, the single particle levels occupied by the last unpaired proton and neutron beyond $N=28$ are assigned by hand as explained above. For $^{48}$K and $^{50}$K, the  ($1f_{7/2},2p_{3/2}$) configurations are used. For $^{52}$K, the configuration ($1f_{7/2},1f_{5/2}$) is used. Now, the theoretical results are in much better agreement with data for both the rapid increase of charge radii and the OES effects. It should be noted that this does not simply imply that the last unpaired proton and neutron really occupy these orbitals. It can also be viewed as a convenient way to choose the deformation of the nuclei studied. Next we check whether further experimental information, such as spin and parity, can help determine which single-particle configurations are preferred.

\subsection{Spin and parity of potassium isotopes}

 For odd-odd nuclei, the spin and parity are  determined by the coupling of last unpaired nucleons, but for odd-even nuclei, the spin and parity are determined by the last unpaired nucleon~\cite{P.Ring,A.Bohr}. In Table~\ref{tab1}, we show the last occupied proton and neutron orbitals of potassium isotopes and the resulting possible spin-parity assignments (fifth column). The deformation parameters $\beta_{20}$ are also shown (seventh column) in comparison with the FRDM results (last column)~\cite{MOLLER20161}. The combinations of $j_{z}$, namely the maximum eigenvalue of the projection of angular momentum $j$ on the $z-$axis, and Nilsson quantum numbers $[N,n_{z},m_{l}]$ are employed to label the orbits occupied by the last unpaired nucleons. Compared to the experimental assignments shown in the sixth column, it is clear that the self-consistent theoretical spin-parity assignments are reasonable. On the other hand, if we chose the configurations fixed by hand as explained above, the theoretical spin-parity assignments (denoted by  blue) do not seem to agree with data.  As a result, only if $^{48,50,52}$K are deformed with $\beta_{20}\approx-0.2$, one could reconcile the experimental measurements and theoretical results for their charge radii.

\begin{table*}\renewcommand{\baselinestretch}{1.5}
\small %
\caption{Spin and parity of  $^{36-52}$K (the fifth column) in comparison with the experimental values (the sixth column)~[12]. The self-consistently determined single particle levels of last unpaired nucleons are listed in the third (proton) and fourth (neutron) columns by the maximum eigenvalue of the projection of angular momentum $j$ on the $z-$axis and the Nilsson quantum numbers $[N,n_{z},m_{l}]$. The manually chosen configurations for $^{48,50,52}$K are noted by blue color. The quadrupole deformation parameters $\beta_{20}$ are shown in the seventh column, in comparison with the FRDM results~[62]. The numbers in red highlight the consistency with the experimental assignments.}\label{tab1}
\begin{tabular}{c|ccccccc}
\hline
A  &  N  & proton  & neutron  &   $j^{\pi}$(this work) &$I^{\pi}$(Exp.)  & $\beta_{20}$(this work)  &  $\beta_{20}$(FRDM)\\
\hline
36    & 17 & ${3/2}^{+}[220]$ & ${3/2}^{+}[202]$ & $0^{+}$, $1^{+}$, $\color{red}{2^{+}}$,  $3^{+}$    &  $2^{+}$ & $-0.12$ & $-0.03$\\
37    & 18 & ${3/2}^{+}[220]$ & ${3/2}^{+}[202]$ & $1/2^{+}$, $\color{red}{3/2^{+}}$  &  $3/2^{+}$ & $-0.14$ & $-0.06$ \\
38    & 19 & ${3/2}^{+}[220]$ & ${3/2}^{+}[220]$ & $0^{+}$, $1^{+}$, $2^{+}$, $\color{red}{3^{+}}$    &  $3^{+}$  & $-0.08$ & $-0.04$\\
39    & 20 & ${3/2}^{+}[211]$ & ${3/2}^{+}[211]$ & $1/2^{+}$, $\color{red}{3/2^{+}}$  &  $3/2^{+}$  & $-0.03$ & $-0.03$\\
40    & 21 & ${3/2}^{+}[220]$ & ${7/2}^{-}[303]$ & $2^{-}$, $3^{-}$, $\color{red}{4^{-}}$, $5^{-}$  &  $4^{-}$  & $-0.08$ & $-0.05$\\
41    & 22 & ${3/2}^{+}[220]$ & ${7/2}^{-}[303]$ & $1/2^{+}$, $\color{red}{3/2^{+}}$  &  $3/2^{+}$  & $-0.07$ & $-0.03$\\
42    & 23 & ${3/2}^{+}[220]$ & ${7/2}^{-}[312]$ & $\color{red}{2^{-}}$, $3^{-}$, $4^{-}$,  $5^{-}$   &  $2^{-}$  & $-0.09$ & $-0.05$\\
43    & 24 & ${3/2}^{+}[220]$ & ${7/2}^{-}[312]$ & $1/2^{+}$, $\color{red}{3/2^{+}}$  &  $3/2^{+}$  & $-0.09$ & $-0.05$\\
44    & 25 & ${3/2}^{+}[220]$ & ${7/2}^{-}[321]$ & $\color{red}{2^{-}}$, $3^{-}$, $4^{-}$,  $5^{-}$ &  $2^{-}$  & $-0.10$ & $-0.06$\\
45    & 26 & ${3/2}^{+}[220]$ & ${7/2}^{-}[321]$ & $1/2^{+}$, $\color{red}{3/2^{+}}$  &  $3/2^{+}$  & $-0.08$ & $-0.05$\\
46    & 27 & ${3/2}^{+}[220]$ & ${7/2}^{-}[310]$ & $\color{red}{2^{-}}$, $3^{-}$, $4^{-}$,  $5^{-}$  &  $2^{-}$  & $-0.08$ & $-0.06$\\
47    & 28 & ${3/2}^{+}[220]$ & ${7/2}^{-}[310]$ & $\color{red}{1/2^{+}}$, $3/2^{+}$  &  $1/2^{+}$ & $-0.00$ & $-0.04$\\
48    & 29 & ${3/2}^{+}[220]$ & ${3/2}^{-}[301]$ & $0^{-}$, $\color{red}{1^{-}}$, $2^{-}$, $3^{-}$ &  $1^{-}$ & $-0.08$ & $-0.05$\\
$\color{blue}{48}^*$    & $\color{blue}{29}$ & $\color{blue}{{7/2}^{-}[303]}$ & $\color{blue}{{3/2}^{-}[301]}$ & $\color{blue}{2^{+}, 3^{+}, 4^{+}, 5^{+}}$ &   & $-0.20$ & \\
49    & 30 & ${3/2}^{-}[220]$ & ${3/2}^{-}[301]$ & $\color{red}{1/2^{+}}$, $3/2^{+}$  &  $1/2^{+}$ & $-0.09$ & $-0.05$\\
50    & 31 & ${3/2}^{+}[220]$ & ${3/2}^{-}[301]$ & $\color{red}{0^{-}}$, $1^{-}$, $2^{-}$, $3^{-}$ &  $0^{-}$ & $-0.12$ & $-0.05$\\
$\color{blue}{50}^*$    & $\color{blue}{31}$ & $\color{blue}{{7/2}^{-}[303]}$ & $\color{blue}{{3/2}^{-}[301]}$ & $\color{blue}{2^{+}, 3^{+}, 4^{+}, 5^{+}}$ &   & $-0.22$ & \\
51    & 32 & ${3/2}^{+}[220]$ & ${3/2}^{-}[301]$ & $1/2^{+}$, $\color{red}{3/2^{+}}$  &  $3/2^{+}$ & $-0.11$ & $-0.08$\\
52    & 33 & ${3/2}^{+}[220]$ & ${5/2}^{-}[310]$ & $1^{-}$, $\color{red}{2^{-}}$, $3^{-}$, $4^{-}$ &  $2^{-}$ & $-0.10$ & $-0.14$\\
$\color{blue}{52}^*$    & $\color{blue}{33}$ & $\color{blue}{{7/2}^{-}[303]}$ & $\color{blue}{{5/2}^{-}[303]}$ & $\color{blue}{1^{+}, 2^{+}, 3^{+}, 4^{+}, 5^{+}, 6^{+}}$ &   & $-0.23$ & \\
\hline
\end{tabular}
\end{table*}

\begin{figure}
\centering
\includegraphics[width=0.8\linewidth]{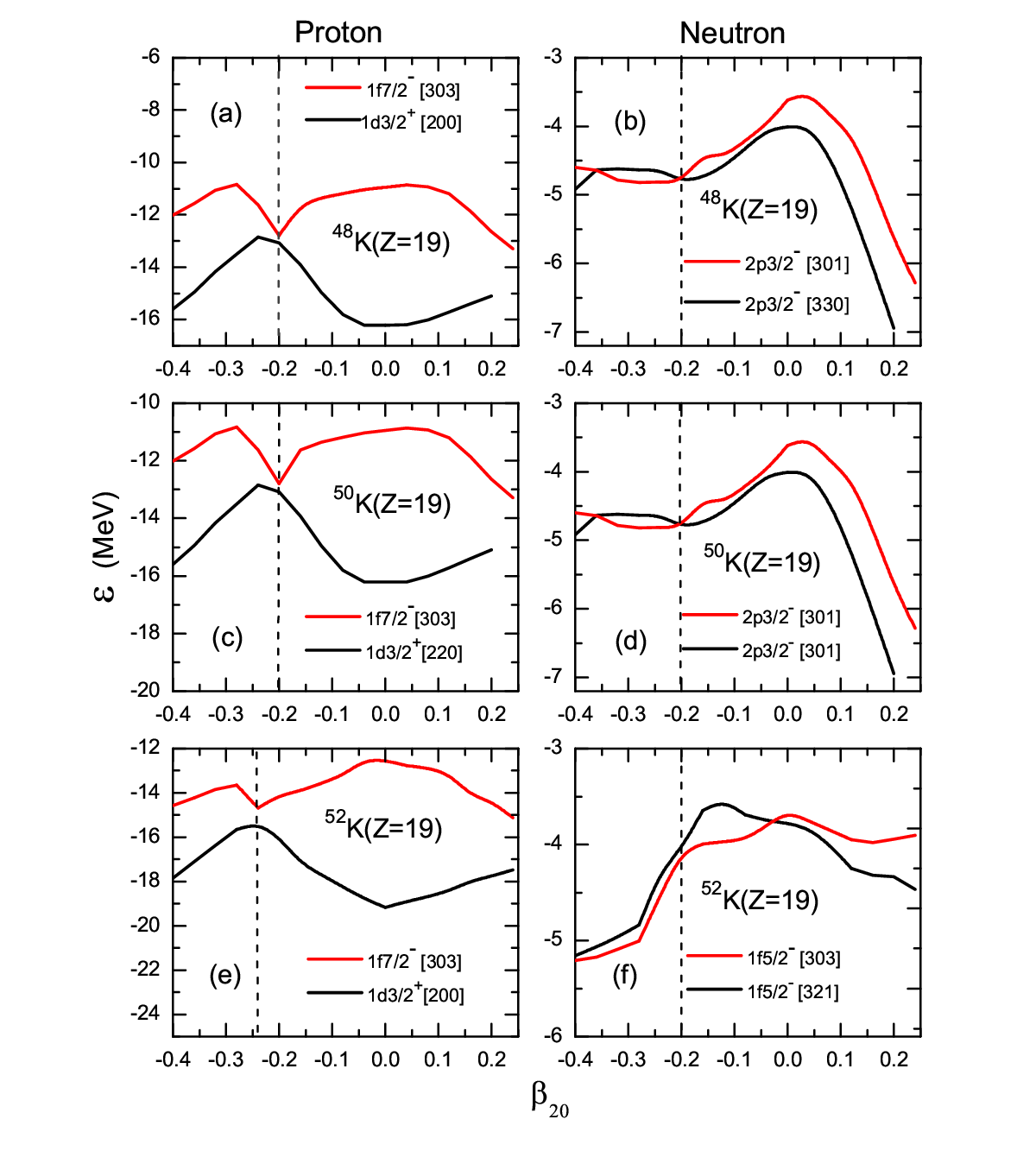}
\caption{Evolution of single particle levels of last unpaired proton (left panel) and neutron (right panel) plotted for $^{48,50,52}$K isotopes. The red-solid lines represent the manually fixed single particle levels  and the  black-solid lines show the self-consistently determined ones.}\label{fig5}
\end{figure}

In Fig.~\ref{fig5}, we show the evolution of single particle (s.p.) levels of last unpaired proton (left panel) and neutron (right panel) for $^{48,50,32}$K. It is clear that around $\beta_{20}\approx-0.2$, the manually fixed s.p. levels and the self-consistently determined ones come closer to each other.

For the self-consistent calculation, if the deformation parameter is restricted to around $\beta_{20}\approx-0.20$, the charge radii $R_{ch}$ of $^{48,50,52}$K are 3.4882 fm, 3.5210 fm and 3.5463 fm, respectively, consistent with the data. This suggests that it is quite likely that the mismatch between the self-consistently determined charge radii and the experimental data is due to the deformation effect, which needs to be studied in more detail in the future.

\section{SUMMARY AND OUTLOOK}
In the present work, we applied the newly proposed RMF(BCS)* ansatz to study the charge radii of the potassium isotopic chain. The parabolic-like shape  between $N=20$ and $N=28$ can be  reproduced very well~\cite{ANGELI201369}, with the odd-even staggerings reproduced as well but slightly overestimated. Beyond $N=28$, the rapid increase of charge radii is also reproduced but now the predicted OES effects are much larger, in contradiction with the experimental data, but in agreement with the Fayans density functional theory.

By carefully studying the impact of different occupation of single particle levels by the unpaired proton and neutron, we found that the overestimated OES effects can be reduced if the last unpaired proton occupies the $1f_{7/2}$ orbital, instead of the self-consistently determined $1d_{3/2}$ orbit. The resulting quadrupole deformation of $^{48,50,52}$K is found to be $\beta_{20}\approx-0.2$. A further study of the experimental spin-parity assignments for these nuclei revealed that, however, the occupation of the $1f_{7/2}$ is not very likely. On the other hand, if these nuclei are deformed instead of spherical, the experimental charge radii data can be reproduced. Judging from the rather soft potential energy surfaces, such an explanation is reasonable and should be checked by beyond-mean-field studies. As a result, we conclude that the latest state-of-the-art experimental measurements of charge radii indeed could help put more constraints on theoretical models.

\begin{acknowledgments}

This work is partly supported by the National Natural Science Foundation of China under Grant Nos. 11735003, 11975041, 11775014 and 11961141004, and the fundamental Research Funds for the Central Universities. This work is also supported by the National Natural Science Foundation of China under Grants No. 11635003, No. 11025524, No. 11161130520, No. 12047513, the Reform and Development Project of Beijing Academy of Science and Technology under Grant No. 13001-2110.

\end{acknowledgments}

\bibliography{refs}
\end{document}